\documentclass[useAMS,usenatbib]{mn2e}
\input psfig.sty

\title[Power spectra including Baryons]{High accuracy power spectra
including baryonic physics in dynamical Dark Energy models}
\author[Casarini \etal]
{L. Casarini$^{1,2}$\thanks{E-mail: casarini@mib.infn.it}, A.V. Macci\`o$^{3}$,
S.A. Bonometto$^{1,2}$, 
G.S. Stinson$^{4}$\\
$^{1}$Department of Physics G.~Occhialini -- Milano--Bicocca
 University, Piazza della Scienza 3, 20126 Milano, Italy \\
$^{2}$I.N.F.N., Sezione di Milano, Milano Italy.\\
$^{3}$Max-Planck-Institut f\"ur Astronomie, K\"onigstuhl 17, 69117 Heidelberg, Germany\\
$^{4}$Jeremiah Horrocks Institute, University of Central Lancashire, Presto PR1 2HE}

\usepackage{graphicx}
\begin{document}

\newcommand{\Nature}{{\it Nature\/} }
\newcommand{\ApJ}{{\it Astrophys. J.\/} }
\newcommand{\ApJS}{{\it Astrophys. J. Suppl.\/} }
\newcommand{\MNRAS}{{\it Mon. Not. R. Astron. Soc.\/} }
\newcommand{\PhRv}{{\it Phys. Rev.\/} }
\newcommand{\PhL}{{\it Phys. Lett.\/} }
\newcommand{\JCAP}{{\it J. Cosmol. Astropart. Phys.\/} }
\newcommand{\AeA}{{\it Astronom. Astrophys.\/} }
\newcommand{\etall}{{\it et al.\/} }
\newcommand{\etal}{et~al.~}
\newcommand{\arXiv}{{\it arXiv: \/} }

\def \apj  {ApJ}
\def \apjl  {ApJL}
\def \apjs  {ApJS}
\def \prd {Phy.Rev.D}
\def \mnras {MNRAS}
\def \hMsun {\ifmmode h^{-1}\,\rm M_{\odot} \else $h^{-1}\,\rm M_{\odot}$ \fi}
\def \hMpc {\ifmmode h^{-1}\,\rm Mpc \else $h^{-1}\,\rm Mpc$ \fi}
\def \Mpch {\ifmmode h\ {\rm Mpc}^{-1} \else $h\ {\rm Mpc}^{-1}$ \fi}
\def \hkpc {\ifmmode h^{-1}\,\rm kpc \else $h^{-1}\,\rm kpc$ \fi}
\def \LCDM {\ifmmode \Lambda{\rm CDM} \else $\Lambda{\rm CDM}$ \fi}

\date{Accepted ............ Received ....................... }

\pagerange{\pageref{firstpage}--\pageref{lastpage}} \pubyear{2002}

\maketitle

\label{firstpage}

\begin{abstract}

The next generation mass probes will obtain information on non--linear
power spectra $P(k,z)$ and their evolution, allowing us to
investigate the nature of Dark Energy. To exploit such data we need high
precision simulations, extending at least up to scales of 
$k \simeq 10\Mpch$, where the effects of baryons can no longer be 
neglected.
In this paper, we present a series of large scale hydrodynamical
simulations for \LCDM and dynamical Dark Energy (dDE) models,
in which the equation of state parameter is $z$-dependent.
The simulations include gas cooling, star formation and Supernovae
feedback.  They closely approximate the observed star formation rate
and the observationally derived star/Dark Matter mass ratio in collapsed systems.
Baryon dynamics cause spectral shifts exceeding $1\, \%$ at $k > 2-3 h$Mpc$^{-1}$ compared to pure n-body 
simulations in the \LCDM simulations.  This agrees with previous studies, although we find a smaller effect 
($\sim 50\%$) on the power spectrum amplitude at higher $k$'s. dDE exhibits similar behaviour, even though 
the dDE simulations produce$\sim 20\, \%$ less stars than the analogous \LCDM cosmologies.
Finally, we show that the technique introduced in Casarini \etal
to obtain spectra for any $w(z)$ cosmology from constant--$w$ models 
at any redshift still holds when gas physics is taken into account.
While this relieves the need to explore the entire functional space 
of dark energy state equations, we illustrate a severe risk that future data 
analysis could lead to misinterpretation of the DE state equation.

\end{abstract}


\section{Introduction}
There can be little doubt that Dark Energy (DE) is a necessary
ingredient for any cosmological model approaching data (see, {\it
e.g.}, the data fits in WMAP5 release, Komatsu \etal 2009). In
turn, the quest for the nature of DE is perhaps the main challenge in
contemporary physics in spite of the fact that no available data conflict
with \LCDM models, in which the DE state parameter is $w \equiv
-1$, as is true for a false vacuum. A false vacuum interpretation of DE, however, leads to an unnatural 
value for its density (fine tuning problem)
and to a number of coincidences characterizing our epoch and the
primeval fluctuation amplitude (coincidence problem).

Many efforts have therefore been devoted to finding viable
alternatives.  Among them, the possibility that DE is a scalar field, $\phi,$
self-interacting through a potential $V(\phi)$ (Wetterich 1989, Ratra
\& Peebles 1989), has been widely explored. In this case, both the ratio
between kinetic and potential energies, $\dot \phi^2/2V(\phi)$ and the DE state parameter
$w(z)$, depend on redshift, $z$. In this paradigm, observational data on $w(z)$ would yield $V(\phi)$.  We will refer to this class of models, as dynamical DE (dDE).

Future large tomographic shear surveys, such as the Euclid project
(see, {\it e.g.}, Refr\'egier \etal 2010), will measure
matter density fluctuations and their evolution from the linear to the
non-linear regime with unprecedented precision, approaching $\approx
1\%$ (Huterer \& Takada 2005). To translate these data into
information on DE, we need to predict power spectra and the
evolution of power spectra with similar precision for a wide set of $w(z)$
laws.  The resulting constraints on $w(z)$ will be a substantial step
toward understanding DE physics (see, {\it e.g.}, Hu \&
Tegmark 1999, Dodelson \& Zhang 2005, Manera \& Mota 2005, Mota 2008, La Vacca \etal 2008, 2009, Kristiansen \etal 2010.).

The evolution of the baryonic component of the Universe is highly uncertain and can affect predictions for lensing observables, manifesting as modified structure growth for a fixed cosmic distance scale (Zentner et al 2008, Hearin \& Zentner 2009).

Collisionless N--body simulations have been extensively used to obtain non-linear
power spectra (e.g. Smith \etal 2003).  These simulations have ignored the effect of baryons, which 
comprise $\sim 15\, \%$ of the matter content of the observed Universe, (Komatsu \etal
2009).  Although baryons make up a minor fraction of the Universal matter content and is distributed 
exactly the same as Cold
Dark Matter (CDM) at the onset of the non-linear growth, the final
distribution of baryons in haloes is significantly different than CDM, because
of its separate later dynamical evolution.

Jing \etal (2006) have shown that, in the spectra of \LCDM models, the
difference between N--body simulations and simulations including
baryon physics exceeds 1\% at $k > 2$--3$\, h{\rm Mpc}^{-1}, $ and
increases to $\sim 10$--20$\, \%$ on smaller scales when $k$
approaches 10$\, h{\rm Mpc}^{-1} $. Rudd \etal (2008) and Guillet \etal (2010)
recently confirmed the need for simulations including baryonic physics, even
though their results differed from Jing \etal (2006).

The discrepant
results originate from the different prescriptions used for star
formation, SN feedback, AGN effects, and possibly, the numerical
approach.  To compare different $w(z)$ laws, these uncertainties need
to be 
under control.  
Using the same prescriptions for all models is essential to
compare their spectra, as we do in this work. Their being tuned to
observations is also a need, when aiming to fit future data. The
prescription used here are tuned on available observations, such
tuning being however carried on within the frame of $\Lambda$CDM
cosmologies. This is to be borne in mind, in view of model dependent
effects that we shall detect, e.g. for what concerns star production
in haloes.
%

Dynamical Dark Energy (dDE) simulations, with a variable state
parameter $w(a)$ deduced from scalar field potentials admitting a
tracker solution, have been performed since 2003
(e.g. Klypin \etal
2003, Linder \& Jenkins 2003, Dolag \etal 2004, Macci\`o \etal 2004,
Solevi \etal 2005).  These have been compared with constant-$w$ simulation outputs. Observables 
considered in these papers, only marginally included
spectra. Recently Francis \etal (2007) have shown how spectral
predictions for constant-$w$ models at $z=0$ can also be used to fit
spectra of cosmologies with a state parameter given by a first
degree polynomial:
\begin{equation}
w(z) = w_0 + w'(1-a)~.
\label{wz}
\end{equation}
More precisely, the spectrum $P(k,0)$, in an {\it assigned} ($\cal A$)
model with state parameter $w(z)$, can be approached by a suitable
{\it auxiliary} constant-$w$ model ($\cal W$) if: (i) in $\cal A$ and
$\cal W$, $\Omega_{b,m,tot}$, $h$ and $\sigma_8$ are equal, (ii) the
constant DE state parameter $w$ of $\cal W$ is tuned so that $\cal W$
and $\cal A$ have the same comoving distance from the Last Scattering
Band (LSB) and $z=0$.  In that case, spectral discrepancies stay $<
1\, \%$, up to $k \sim 2$--$3\, h\, $Mpc$^{-1}$ (symbols here above
have their usual meaning).

Francis \etal (2007) also tried extending their results to higher redshifts, but
discrepancies between $\cal A$ and $\cal W$ increased to
several percent; not enough to exploit forthcoming data.
The required precision was, however, recovered through
a new technique introduced by Casarini, Macci\`o \& Bonometto (2009, Paper
I hereafter) and tested with N--body simulations.

The final aim of this work is to extend the techniques use in Paper I to
scales that only hydrodynamical simulation can test. This is
a serious challenge for the description of physical processes
included in hydrodynamical algorithms.  We will devote a part of our work to understanding these challenges.

The focus of this paper is twofold: (i) Study the
importance of the effects of baryons on the matter power spectrum on
the small scales that will probed by future surveys, bearing in mind that the physical
assumptions will have to be carefully calibrated to approach the
unprecedented precision requirements. (ii) Extend the 
commonalities found in $N$-body simulations in Paper I among spectra of different models down to scales where baryon
physics cannot be disregarded. Accordingly, we shall use the same baryon physics in a series of 
simulations to compare the spectra of different dDE models.

The paper is therefore organized as follows: In \S 2, we
describe the approach used in Paper I. In \S 3, we motivate the choice of the
models considered. \S 4 is devoted to describing our
simulations and the techniques used to analyse them. In \S 5, we
present our hydrodynamical simulations.
\S 6 is subdivided in subsections that concern various
aspects of the power spectra analysis. In \S 6 we test: (i) 
the relation between N-body and hydrodynamical spectra (ii)
the effect of box size and numerical resolution (iii)
the level at which model parameters can be discriminated; and, finally, (iv) the success of
the extension of Paper I approach to the higher $k$ range. A general
discussion of the results closes our paper.

\begin{figure}
\psfig{figure=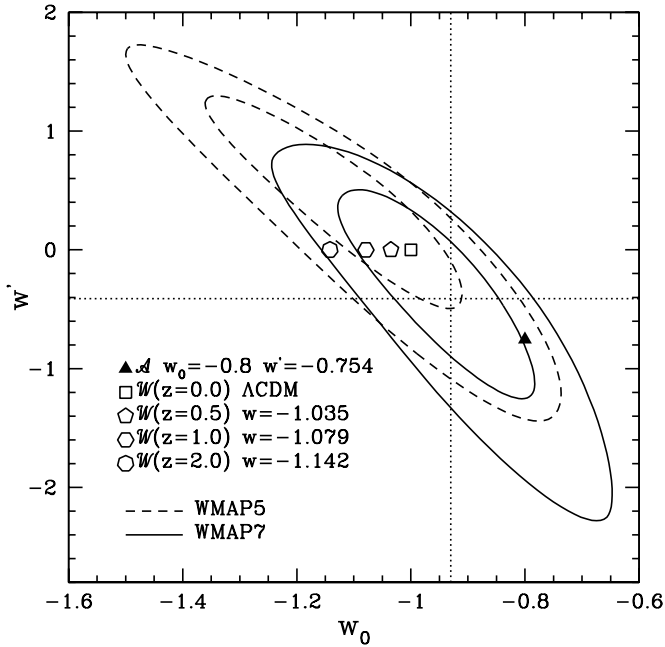,width=0.52\textwidth}
\caption{WMAP7 (solid curves) \& WMAP5 (dashed curves) likelihood
ellipses on the $w_o$--$w'$ plane and the position of our models on
that plane. The significant shift from WMAP5 to WMAP7 is not due to
new CMB data, but mostly from the improved HST determination
of $H_0$. Dotted lines yield the best likelihood model. The selected model
lies near the 1--$\sigma$ contour for WMAP7,
and within 2--$\sigma$'s, for WMAP5 as well. Notice how much closer the $\cal W$ models 
(empty polygons) are than the assumed cosmology (filled triangle).
This highlights a serious observational risk (see \S \ref{sec:dis}).
}
\label{ellypse}
\end{figure}

\section{Approaching dynamical DE spectra}

Here is a short summary of the technique presented in Paper I, which
improved the results of Francis \etal (2007) at $z>0$. Given an
assigned model $\cal A$, we introduce an auxiliary model $\cal
W$$(z)$, for each $z$.  $\cal A$ and $\cal W$$(z)$ are required to
share the values of both $\omega_{b,c,m} = \Omega_{b,c,m} h^2$
and~$\sigma_8$, at the selected $z~.$

The first condition is satisfied at any redshift using
the relationship between $H$ and critical density $\rho_{cr}$ 
\begin{equation}
H^2 = (8\pi G/3) \rho_{cr}~,
\end{equation}
where $H$ is the Hubble parameter. Multiplying both sides of this
equation by $\Omega_{m}$ (or $\Omega_b,~\Omega_c$) produces
\begin{equation}
\Omega_m H^2 = (8\pi G/3) \rho_{m}~.
\end{equation}
The r.h.s. of this equation scales as $a^{-3}$, independent of model, since $\omega_m
\propto \Omega_m H^2$ (or $\omega_b,~\omega_c$). Therefore, once
the $\cal A$ and $\cal W$ models share $\omega_{b,c,m}$ at $z=0$,
they also share $\omega_{b,c,m}$ at every other $z$. Accordingly, each $\cal W$$(z)$ model 
shares the same $\omega_{b,c,m}$ values.

Conversely, the evolution of $\sigma_8$ depends on the state
equation of DE and its value at $z=0$ and can be obtained 
only from the {\it constant} DE state parameters $w(z)$, which is fixed by 
requiring that $\cal W$$(z)$ and $\cal A$ have equal distances between $z$ and the LSB.

The choice of $H_0$ (the Hubble parameter at $z=0$) is 
unconstrained, so we set it equal to $H_0$ in $\cal A$ to allow the simulated volumes to
have the same side $L$ in both Mpc and $h^{-1}$Mpc units. Notice that
a simple-minded generalization of Francis et al.~(2007) criterion to
high $z$ would require equal $\Omega_{b,c,m}(z)$ and, thence, $H(z)$,
and creates serious problems of sample variance and model comparison,
at most redshift values.

In Paper I, the mapping between $\cal W$$(z)$ and $\cal A$ spectra was
tested, using N--body simulations, up to $k \simeq 2$--$3\, h\,
$Mpc$^{-1}$.  The discrepancies were mostly in the per-mil range. The techniques used in Paper 
I worked better for at higher $z$ values.

\section{Cosmological Models}

To extend our results to hydrodynamical simulations, we use three cosmological models, each of 
which is consistent with both WMAP5 and WMAP7 data.  Two models share most cosmological parameters:
baryon and CDM density $\Omega_b = 0.046, ~\Omega_c= 0.228$, Hubble parameter (in
units of 100 km/s/Mpc) $h = 0.7$, r.m.s matter fluctuation amplitude
at $z=0$ on 8~$h^{-1}$Mpc scales,$~\sigma_8 = 0.81$, and primeval
spectral index $n=0.96$.  The models differ in the assumed dark energy equation of state,  
either $w \equiv -1$ for \LCDM or using equation (\ref{wz}) with $ w_o = -0.8, ~w' = -0.754 \,$ 
for dDE (also sometimes referred to as the $\cal A$ model).  We compare the dDE with various 
{\it auxiliary} constant-$w$
models $\cal W$$(z)$. We plan to extend the tests performed on dDE to
other $w(z)$ expressions in future work, but there is currently no reason to believe that 
our dDE model is peculiar. In Figure \ref{ellypse}, the $\cal A$ model and the
auxiliary models are set on the WMAP likelihood ellipses on the
$w_o$--$w'$ plane. 

In order to extend the model comparison, we also considered another
\LCDM model ($\Lambda$CDM1), with a greater CDM density
$\Omega_c = 0.254$, which yields a matter density parameter $\Omega_m =
0.3$.  Thus, $\Lambda$CDM1 is $\sim
2~\sigma$'s from $\Lambda$CDM using the WMAP5 data.  \LCDM is run in boxes of different 
sizes to test the effectiveness 
of the mass and spatial resolution used. A summary of models and boxes 
used in this paper is presented in Table \ref{tab:sims}. \LCDM coincides with $\cal W$$(z=0)$.

\section{Simulations and their analysis}

\begin{table}
 \centering
 \begin{minipage}{130mm}
  \caption{SIMULATION PARAMETERS}
  \begin{tabular}{lccccc}
\hline  & 100 $\Omega_b$ &  $\Omega_c$ & $\sigma_8$  & L [\hMpc] & $z_{\rm end}$ \\
\hline
\LCDM  & 4.6 & 0.228  & 0.81 & 256 & 0 \\
$\Lambda$CDM$_1$  & 4.6 & 0.254  & 0.81 & 256 & 0 \\
$\Lambda$CDM$_{\rm L=64}$  & 4.6 & 0.228  & 0.81 & 64 & 0 \\
dDE ($\cal A$)  & 4.6 & 0.228  & 0.81 & 256 & 0 \\
$(w_o=-.8\, ,$& &  &  &  &  \\ 
$~w'=-.754)$& &  &  &  &  \\ 
\hline
$\cal W$$_{0.5}$& 4.6 & 0.228  & 0.816 & 256 & 0.5 \\ 
(w=--1.035) & &  &  &  &  \\ 
$\cal W$$_{1}$& 4.6 & 0.228  & 0.823 & 256 & 1 \\
(w=--1.079) & &  &  &  &  \\
$\cal W$$_2$& 4.6 & 0.228  & 0.834 & 256 & 2 \\
(w=--1.142) & &  &  &  &  \\
\hline 
\label{tab:sims}
\end{tabular}
\end{minipage}
\end{table}

We run both N-body and hydrodynamical simulations (here below also
dubbed DMO and DMG, representing ``DM-only'' and ``DM \& gas'' respectively). For DMO
simulations we use {\sc pkdgrav} (Stadel 2001) that has been modified to
enable variable $w(z)$ as described in Paper I.  For the hydrodynamical
simulations, we use {\sc gasoline}, a multi-stepping, parallel
Tree\-smoothed particle hydrodynamic (SPH) $N$-body code (Wadsley et al. 2004).

{\sc gasoline} includes radiative and Compton cooling for a primordial mixture of
hydrogen and helium. The star formation algorithm is based on a Jeans
instability criteria (Katz 1992), but simplified so that gas particles satisfying constant 
density and temperature thresholds in convergent flows spawn star particles at a
rate proportional to the local dynamical time (see Stinson et
al. 2006). The star formation efficiency was set to $0.05$ based on simulations of the 
Milky Way that satisfied the Kennicutt (1998) Schmidt Law. The code also includes supernova feedback
as described by Stinson et al. (2006), and a UV background following
Haardt \& Madau (1996); see Governato et al. (2007) for a more
detailed description of the code.
We applied the same modifications to {\sc gasoline} that we implemented in {\sc
pkdgrav} in Paper I to handle the dDE models.

All of the simulations use 256$^3$ CDM particles (and $256^3$ gas particles)
and most of them are boxes 256$\, h^{-1}$Mpc on a side. For the $\Lambda$CDM, dDE models, 
and the related auxiliary models, the CDM particles each have a mass of $m_c
h/M_\odot = 7.61 \times 10^{10}$ in the DMO simulations. In the DMG simulations, the  CDM 
particles have a mass of $m_c h/M_\odot = 6.33 \times 10^{10}$ and the gas particles have 
a mass of $m_b h/M_\odot = 1.28 \times 10^{10}$. For the $\Lambda$CDM1 model, $m_c h/M_\odot
= 8.33 \times 10^{10}$ in the DMO and $m_c h/M_\odot = 7.05 \times 10^{10}$,
$m_b h/M_\odot = 1.28 \times 10^{10}$ in the DMG.

The force resolution (softening) is 1/40 of the intra-particle
separation.  For $L =256\, h^{-1}$Mpc, this corresponds to a distance
$\epsilon \simeq 25~h^{-1}$kpc (wavenumber $\kappa = 2\pi/\epsilon
\simeq 150\, h\, $Mpc$^{-1} $). This softening provides spectral resolution up
 to 
$k=10 h$Mpc$^{-1}$.  Each simulation is started at 
$z_{in} = 24$ 
using the same random realization of the density field.  
In order to have equal $\sigma_8$ at the various redshift $z_i$,
where model spectra are compared, different models are started with
(slightly) different $\sigma_R$ at $z_{in}$. (Here $\sigma_R$ is the
m.s. fluctuation over the scale $R$; e.g.: $\sigma_8 \equiv
\sigma_{8\, h^{-1}{\rm Mpc}}$.) The smallest resolved scale $R_r$
yields the greatest relevant $\sigma_{R_r}$.  The selected $z_{in}$ is
such that residual non--linear effects cannot induce discrepancies
among models, over the scale $R_r$, exceeding 1:10$^4$ (Crocce \etal 2006).
For the \LCDM model (DMO and DMG) we also ran a smaller box 
$L=64\, h^{-1}$Mpc ($\Lambda$CDM-small), to examine the resolution
effects.

\begin{figure}
\psfig{figure=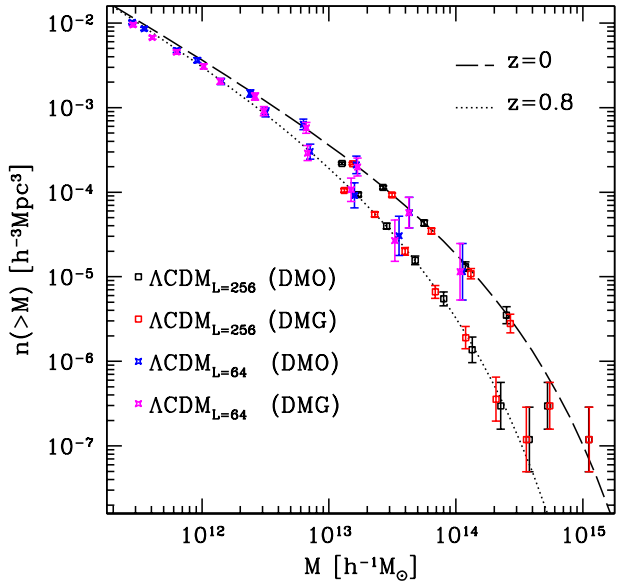,width=0.5\textwidth}
\caption{Mass functions for the $\Lambda$CDM model at two different
redshifts ($z=0; 0.8$) in the N-body and Hydrodynamical simulations: DMO and
DMG respectively. Results for the $L=64 \hMpc$ and $L=256 \hMpc$ boxes
are shown with different symbols. Dashed and dotted lines show the
Sheth \& Tormen (2002) prediction for $z=0$ and $z=0.8$.}
\label{mf}
\end{figure}
One of the key analyses in this work are power spectra of 
the matter density field.  The matter density field is found by interpolating 
the particle distribution onto a regular $N_g \times N_g \times N_g$ grid using the 
Cloud-in-Cell algorithm.   The power spectra are calculated using a FFT (Fast Fourier Transform) 
of the matter density field. For our analysis we use  $N_g=2048$ for the large boxes
($L=256 \hMpc$) and $N_g=512$ for \LCDM-small, in order to cut the spectra at the same
frequency.

Another key piece of analysis was finding haloes.  We identified collapsed structures 
using the spherical overdensity (SO) algorithm.  We used a time varying virial density
contrast from the fitting formulae of Mainini \etal
(2003).  The halo catalogue includes all structures with more
than 200 DM particles.  For DMG simulations we estimated the halo
masses by taking into account {\it all} particles inside the virial
radius (see Macci\`o \etal 2008 for further details on our halo
finding algorithm).

\section{Mass functions and star formation}
\label{mfandsf}
\begin{figure}
\psfig{figure=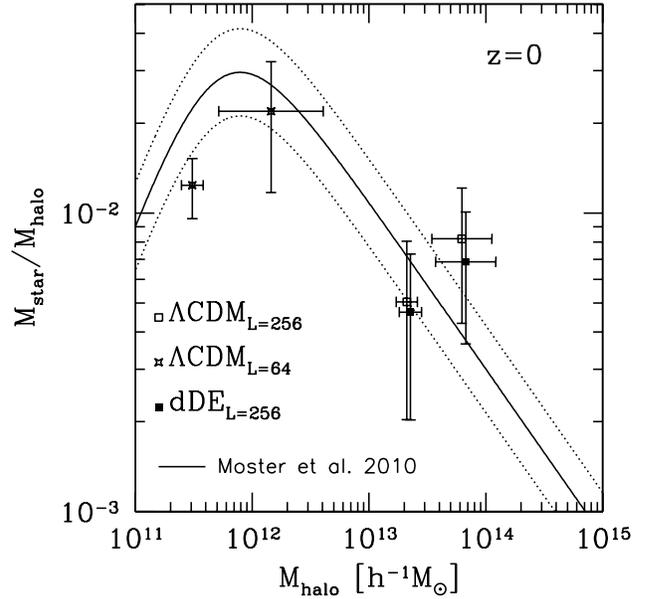,width=0.5\textwidth}
\caption{Ratio between stellar mass and total mass in haloes.
Empty and solid squares show the results for the large box ($L=256$)
in the \LCDM and dDE models. Crosses show results for the \LCDM
small box ($L=64$).  The solid line represents the recent
observational results from Moster \etal (2010), dotted lines show
a 40\% scatter ($\sim 2\sigma$) around the mean. Notice that dDE
yields a smaller $M_{star}$, at a fixed $M_{halo}$ mass.  The slightly
overproduction of stars at high halo masses is due to the absence of
AGN-like feedback in our hydrodynamical simulations.}
\label{starhalo}
\end{figure}

\begin{figure}
\psfig{figure=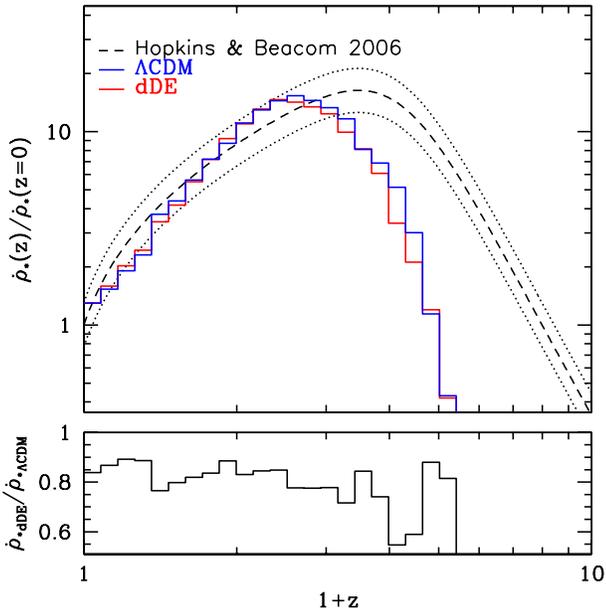,width=0.5\textwidth}
\caption{Star formation rate in DMG simulation. The upper panel shows
the evolution of the star production rate {\it vs.}~$z$: solid
histograms are \LCDM and dDE rates (blue and red respectively) in the
256$\, h^{-1}$Mpc box; the dashed (black) line is Hopkins \& Beacom
fit (2006, assuming a Salpeter IMF); all curves are normalized to the
$z=0$ value. The apparent high--$z$ discrepancy is due to the
low--mass halo cutoff ($\approx 200\, m_c \simeq 1.3 \times 10^{12} h^{-1}
M_\odot$) in the simulations.  The lower panel shows the ratio
between star productions in dDE and \LCDM models. The former model
clearly exhibits a lower stellar production at all redshifts.}
\label{sfr}
\end{figure}

A significant test of our simulations are halo mass functions. 
Figure \ref{mf} compares the cumulative mass function
$N(>M)$ in our \LCDM simulations with the Sheth \& Tormen (2002)
predictions for $z=0$ and $z=1$.  Figure \ref{mf} shows the haloes found in the DMO and DMG simulations for both 256 and 64$\, h^{-1}$Mpc boxes.  The consistency between
the DMO and DMG simulations is readily apparent, as is the continuity of the mass function
over 4 orders of magnitude.

We make a more direct comparison with observations in Figure \ref{starhalo}, which shows 
how the ratio $M_{star}/M_{halo}$ depends on the halo mass in both 64 and 256$\, h^{-1}$Mpc 
boxes. Figure \ref{starhalo} shows the average $M_{star}/M_{halo}$ for each mass bin.  
These are compared with the predictions from the halo occupation model of Moster \etal
(2010). Each simulation point is obtained from equal halo
numbers and horizontal error bars yield the m.s. spread of
$\log(M_{halo})$.  Even though the simulations are run at low resolution relative to star 
formation scales, the number of stars formed in the haloes generally follows the halo 
occupation distribution trend.  It is possible that too many stars form in haloes in 
the highest $M_{halo}$ bin.  This is likely due to overcooling that plagues hydrodynamic 
simulations and might be fixed by including AGN
effects in the simulations. 

Most important to the current study, Figure \ref{starhalo} shows a significant reduction 
of $M_{star}/M_{halo}$ from the \LCDM simulations to those run with dDE.  
Such reduction is significant and one should not be
misled by the apparent consistency within error bars: they are large
because of the spread of $M_{star}/M_{halo}$ through individual
haloes, but the reduction of $M_{star}/M_{halo}$ is not randomly
distributed, exhibiting just a slight increasing trend from smaller to
greater haloes. We shall return on this point below.

Let us however remind that star formation in hydrodynamical
simulations strongly depends on numerical (mainly mass) resolution
(e.g. Mayer \etal 2008 and references therein). We therefore expect
the result shown in Figure \ref{starhalo} to change if the same
simulations are run at higher resolution (not only because we would
resolve lower mass haloes). Naively one could expect to see
a higher stellar production and, therefore,
a significant deviation of simulated curves from observed ones.
The point of our comparison with real data is not to show that
our resolution allows intrinsically correct results, but that, {\it
with resolution adopted in this specific work}, hydro simulations produce a 
reasonable amount of stars.

Figure \ref{sfr} shows the cosmic star formation history in the 256$\, h^{-1}$Mpc 
boxes for both the \LCDM and dDE simulations.  The top panel shows the simulated star 
formation rate density (SFR, $\dot \rho_*$) 
normalized to the observed $z=0$ value from 
Hopkins \& Beacon (2006).  Due to the low resolution of the simulations, they underproduced 
stars by a factor of 
$\approx 1.5$.  We renormalized these values to show that the shape of the star formation 
history is similar to observations.  The observational results are shown with constant $\pm 20\, \%$ 
error bars that roughly approximate 
the 2-$\sigma$ uncertainty of their results.
Besides the slight deficiency in star formation, there are two key differences between 
the simulations and observed cosmic SFH.  The Hopkins \& Beacon (2006) observations peak at $z \simeq 2.3\, $
while the simulations peak around $z \simeq 1.5$. The onset of star formation is also later 
in the simulations than the observations.  This discrepancy is again a result of low resolution 
in the 256$\, h^{-1}$Mpc boxes.  Christensen \etal (2010) showed that stars don't form in halos 
with less than 200 gas particles.  Such haloes have a mass of $\sim 1.3 \times 10^{12}\, h^{-1} 
M_\odot$ in our simulations.  Sufficient numbers of these haloes do not form in our simulations 
until $z=4$ to form stars.  The SFH of the 64$\, h^{-1}$Mpc box more closely follows the observed 
SFH with a peak around $z \simeq 2.5$ and an early onset of star formation.  The results in this
case are also affected by significant sample variance due
to the limited size of the simulated volume.

The bottom frame of Figure \ref{sfr} shows the ratio between $\dot
\rho_*$ the dDE and \LCDM simulations, highlighting the reduced star
formation found using the dDE model.  The decrease is $\sim 20 \%$ and
it is worth noting that it originates simply from a change of the DE
equation of state.  The models, in fact, have exactly the same age at
$z=0$ and identical cosmological parameters, apart from $w$.  The
$\sim 20\, \%$ discrepancy shows how different the evolutionary
histories and concentration distributions are between the two
cosmologies.  The magnitude of the discrepancy is larger than
anticipated and we will devote future work to determining the cause of
such a significant difference.

Before concluding this section let us however return on the issue
concerning resolution. The comparison with real data just shows that
our hydro simulations give a reasonable star production rate, as
function of halo mass, and a reasonable distribution of such rate, as
a function of redshift. This ensures us that the effect of baryonic
matter (mainly stars) on the total matter power spectrum should be
realistic. This will be true especially on the spatial scales ($R$) 
that we will address in this work (for $k\approx 10 \rightarrow 
R\approx 300 \hkpc $), scales that are much larger 
than the typical sizes of galaxies.

\section{Spectra}

Now that we have shown that our hydrodynamic model is reasonable on the large scales we wish to study, 
we turn our attention to how the density power spectra depend on the DE model used.

\subsection{Hydro vs.~N--body simulations}

\begin{figure}
\psfig{figure=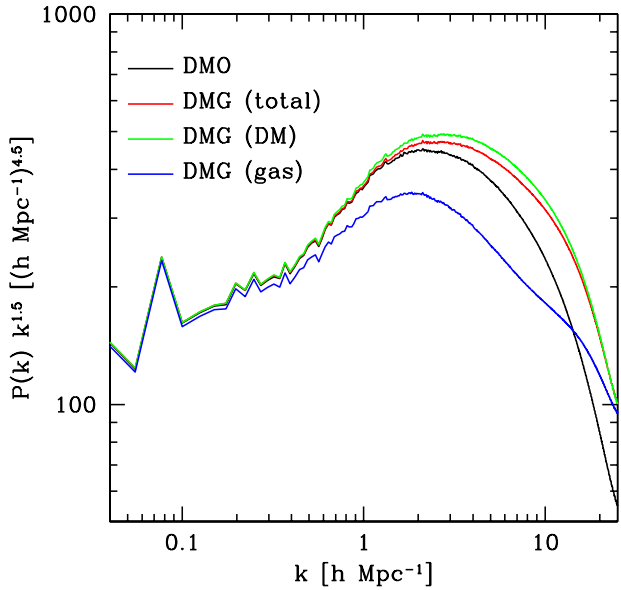,width=0.5\textwidth}
\caption{Spectra for different components in the  $\Lambda$CDM  256$\, h^{-1}$Mpc box. 
Results for dark matter only (DMO) and dark matter \& gas (DMG) components
are shown using different colors.}
\label{lw}
\end{figure}

Figure \ref{lw} shows a comparison between DMO and DMG spectra 
at $z=0$. Both the DM and total power spectra of the collisional case show a substantial enhancement especially at scales
$ k> 2 \Mpch$. This scale corresponds to the most massive structure in our simulation
and reflects the fact that gas has condensed and cooled inside those structures driving 
a contraction of the haloes themselves (e.g. Gnedin \etal 2004).
The gas shows a bias with respect to the other DMG spectra at large scales $k< 8-10 \Mpch$. 
This effect has been shown in previous work (Jing \etal 2006, Rudd \etal 2008)
and is related to gas moving from large scales to small scales due to cooling.
The cooling is most obvious at $k>10 \Mpch$, where the gas spectra has more power than DMO,
reflecting the presence of dense clumps of cold gas at small scales, in the center of dark matter halos.

\begin{figure}
\psfig{figure=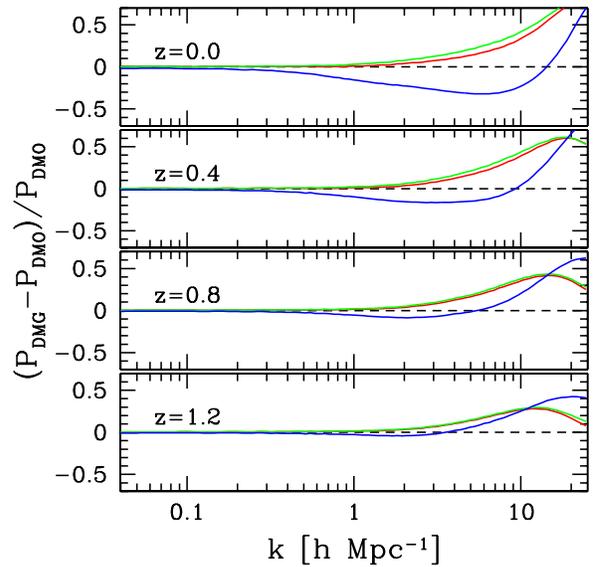,width=0.5\textwidth}
\caption{DMG {\it vs.}~DMO spectra comparison for the 256 \hMpc box at different redshifts.}
\label{big}
\end{figure}

The differences between DMO and DMG can be better appreciated in Figure \ref{big} where the ratio 
between the different spectra is shown as a function of redshift.
It is interesting to note that the large scale bias of the gas spectrum already 
shown in Figure \ref{lw} is strongly 
redshift dependent and starts to develop only after $z \approx 1$ and is related to gas cooling and consumption within
dark matter haloes. Another interesting feature at $z=0$ is the higher power in the DM spectrum with respect 
to the total one in the DMG simulation at intermediate scales ($2<k/(\hMpc)<10$).
This is due to a combination of two effects: the large bias in the gas spectrum 
(which reduces the total one) and the small contribution of the stellar component
on these scales, that are larger than typical galactic scales.
For $k>10 \Mpch$ the gap between the total and DM spectra vanishes thanks to the increased 
contribution of cold gas and stars. We will come back to discuss the effect of stars and resolution on 
DMG spectra in section \ref{ssec:res}

\begin{figure}
\psfig{figure=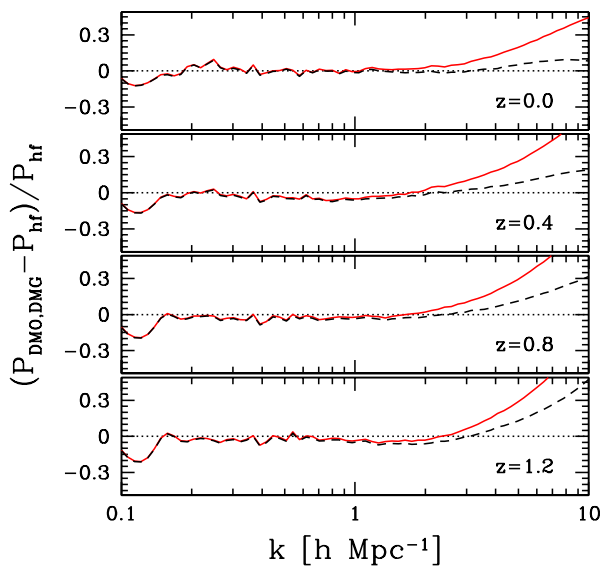,width=0.5\textwidth}
\caption{Comparison between spectra from simulations and halofit
predictions for the \LCDM model in the $L=256 \hMpc$ box. The dashed
(black) curve shows the ratio between matter spectrum in the DMO
simulation and halofit. The solid (red) curve shows the same ratio but
for the total (dark+star+gas) spectrum in the DMG simulation.}
\label{halofit}
\end{figure}

A further interesting result is the comparison between DMO or DMG
spectra and the {\sc halofit} expressions (Smith \etal 2003). Only the 256$\, h^{-1}$Mpc box
can be used for this comparison.  The results are shown in Figure
\ref{halofit}. The validity of {\sc halofit} up to $k < 2$--$3\, h\,
$Mpc$^{-1}$ is confirmed, although discrepancies $\cal O$$(6\, \%)$
are found in some spectral ranges. A clear conclusion is that, when
aiming at 1$\, \%$ precision, the {\sc halofit} expressions need
to be significantly improved.

\subsection{Box size and resolution effects}
\label{ssec:res}

\begin{figure}
\psfig{figure=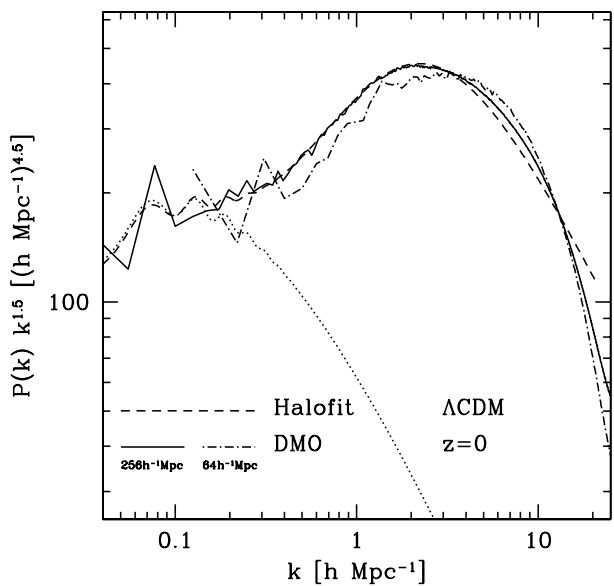,width=0.5\textwidth}
\caption{Comparison between DMO spectra in the large ($L=256 \hMpc$,
solid line) and small ($L=64 \hMpc$, dotted-dashed line) boxes {\it
vs.}  {\sc halofit} expressions for the $\Lambda$CDM cosmology. The
large box is in reasonable agreement with {\sc halofit} in its
validity range ($k<3 h \rm Mpc^{-1}$); while the small box shows a
shortage of power in that range, due to the lack of long wavelength
components.}
\label{whynbody}
\end{figure}

Figure \ref{whynbody} compares the power spectra of the
large and small DMO boxes with the {\sc halofit} expressions at $z=0$.
While the power spectrum from the 256$\, h^{-1}$Mpc box
agrees with {\sc halofit} across the entire range of its validity, the spectrum
from the 64$\, h^{-1}$Mpc box exhibits a shortage of power in the
spectral region where non-linearity begins to display its effects.
This shortage of power is due to the lack of long wavelength components in the small box, 
and clearly shows that $L \approx 60$--$70\, h^{-1}$Mpc boxes are effected by
significant sample variance.  Thus, we cannot draw quantitative conclusions above $k \simeq
0.5$--$1\, h\, $Mpc$^{-1}$ from such small boxes.
Conversely hydrodynamical simulation results could be dependent on 
numerical resolution, and the higher resolution possible in small boxes 
make them a useful tool to address resolution issues.

\begin{figure}
\psfig{figure=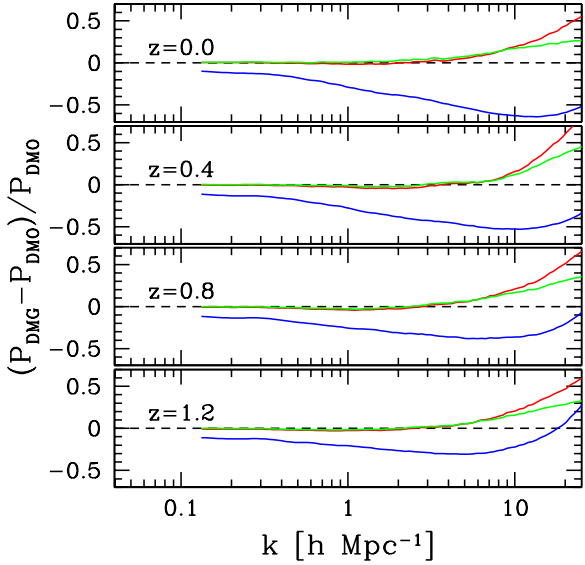,width=0.5\textwidth}
\caption{DMG {\it vs.}~DMO spectra comparison for the small box ($L=256 \hMpc$) at 
different redshifts.}
\label{small}
\end{figure}

Figure \ref{small} shows the fractional difference between DMO and DMG simulations
in the small box. The gas component shows a slightly different behaviour than it did in 
Figure \ref{big}. In the small box, the bias at large scales is already present at $z=1.2$ and
grows substantially to $z=0$, propagating its effects at smaller and smaller scales.
This is a consequence of the enhanced SF in the small box that, as mentioned in \S \ref{mfandsf}, peaks at $z=2.4$. 
Owing to the (smaller) discrepancies in the overall spectra, it seems
natural to argue that a systematic power depletion in the gas spectrum
occurs because of stars formation in small scale sites.
This is confirmed by a comparison with Rudd \etal (2008) spectra,
shown in their Figure 2, right-hand panels. In their case, the power
depletion in the gas spectra is even stronger, so that the overall spectra
are systematically above CDM. 
However, Rudd \etal (2008) state in their paper that their simulations are
characterized by an excess in star formation by a factor $\sim 3$.
Because of the high star formation, they see the same depletion in their gas power spectra, only stronger.
The enhanced star formation in the small box also has an effect on the ratio
between total and DM spectrum in the DMG simulation. In the small box, the total spectrum 
is higher than the DM spectrum for $k>10 \Mpch$ thanks to the stellar component.

The effect more important than resolution is the intrinsic lack of power in the small box.
Thus, the total spectrum is strongly 
reduced in the small box with respect to the large one. This difference 
is on the order of 50\% at $k=10 \Mpch$ and even larger, up to 100\% for 
$k=20 \Mpch$ (see also Levine \& Gnedin 2006).
It is interesting to note that our results on the total, gas and DM spectra behaviour 
in the large box are closest to the results of Jing \etal 2006  (who used a box
of 100 \hMpc), who also found the DM spectrum to lie above the total one for intermediate 
values of $k$.

The two conclusions we draw from this comparison between the small and large
box are:  (i) $L \sim 60$--$70\, h^{-1}$Mpc boxes are effected by
significant sample variance, frequently yielding potentially
misleading results, and hardly allowing quantitative conclusions above $k
\simeq 0.5$--$1\, h\, $Mpc$^{-1}$; (ii) in turn, boxes with a side $L \sim
250$--260$\, h^{-1}$Mpc yield results that will only require slight quantitative
adjustments (possibly at higher resolution) for $k > 6$--$7\, h\, $Mpc$^{-1}$.

We are thus left with no reasons why we should not trust spectral
comparisons between different models, shown in the forthcoming
Figures. If some (minor) bias is present, it should affect related
spectra in a similar fashion.

%

\subsection{dDE and spectral regularities}
We now turn our attention from the effects of resolution and baryons on power spectra 
to the effects of using different cosmologies.

\begin{figure}
\psfig{figure=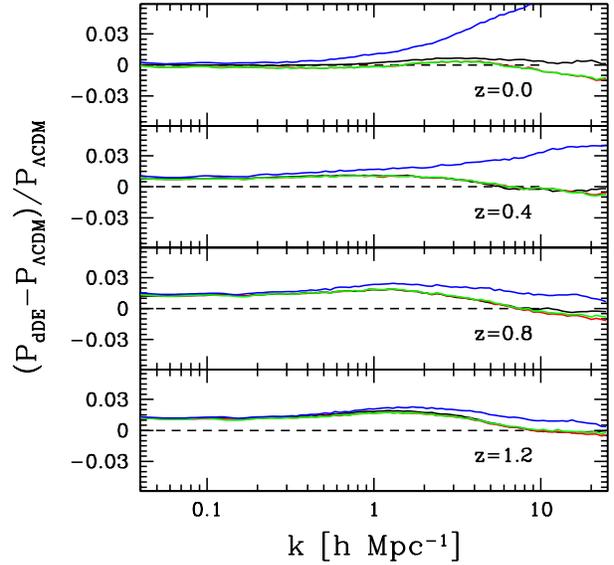,width=0.5\textwidth}
\psfig{figure=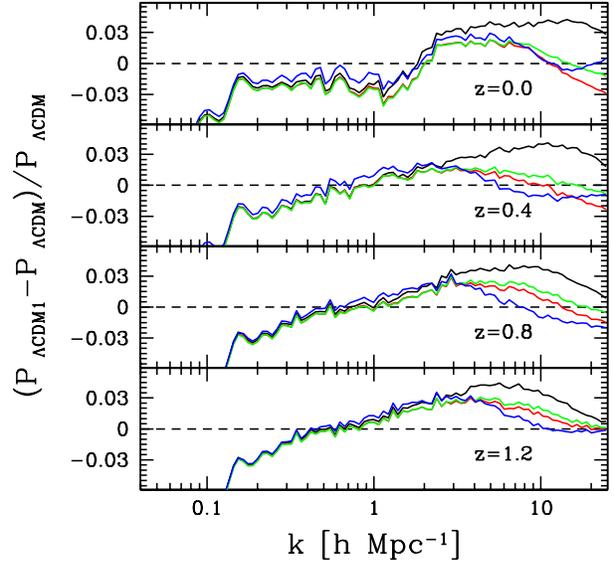,width=0.5\textwidth}
\caption{Fractional spectral differences comparisons: \LCDM {\it vs.}~dDE  is in the upper
plot and \LCDM~{\it vs.}~\LCDM1 is in the lower plot in DMO and DMG
simulations.  Four different redshift values are shown. Color
meaning are the same as in Figure \ref{lw}.}
\label{dDE}
\end{figure}

Figure \ref{dDE} shows the fractional spectral differences between 
\LCDM and dDE (upper panel) and \LCDM and \LCDM1 (lower panel).
Two main effects are apparent: (i) the curves are much
smoother for dDE than for \LCDM1. In particular, in the top panel, the
discrepancy at $z=0$ almost vanishes, for the total spectrum.  This is because 
\LCDM is the $\cal W$$(z=0)$ model of dDE, so these
findings are expected. In particular, the discrepancies up to $\sim
3\, \%$ at higher $z$ affirm the findings of Francis \etal (2007).
The discrepancies between \LCDM and \LCDM1 (bottom
panel) are far more exaggerated and are typical of models lying $\sim 2 \sigma$'s away from
WMAP5 data.  This shows that spectral analysis that can detect 1$\, \%$
differences will successfully discriminate among models within the
2-$\sigma$ curves in Figure \ref{ellypse}. (ii) The $\Lambda$CDM1 model
re-affirms that DMO--DMG spectral discrepancies become most significant at $k \simeq
2$--3$\, h$Mpc$^{-1}$.

In Section 2 we reviewed the technique presented in Paper I, that we
shall explicitly test here. We start from the assigned model $\cal A$,
which is our dDE model, and find the auxiliary models $\cal W$$(z)$
for at $z = 0$, 1 and 2~. They must share the values of
$\omega_{b,c,m} = \Omega_{b,c,m} h^2$ and~$\sigma_8$ with $\cal A$.
Let us recall that the former request is easily fulfilled, as $ H^2 =
(8\pi G/3) \rho_{cr} $ and, by multiplying both sides of this relation
by $\Omega_{m}$ (or $\Omega_b,~\Omega_c$) we have that $ \omega_m
\propto \rho_{m} \propto a^{-3}$, independent of the model.
Conversely, the evolution of $\sigma_8$ depends on the state
equation of DE and its values at $z=0$ and $z=24$ can only be worked out
once we know the {\it constant} DE state parameters $w(z)$ of the
$\cal W$$(z)$ models for $z = 0$, 0.5, 1, 2~.
The final requirement, causing the dependence on $z$ of the constant
$w$'s is that $w$ is tuned so that $\cal W$$(z)$ and $\cal A$ have equal
distances between $z$ and the LSB.

\begin{figure}
\psfig{figure=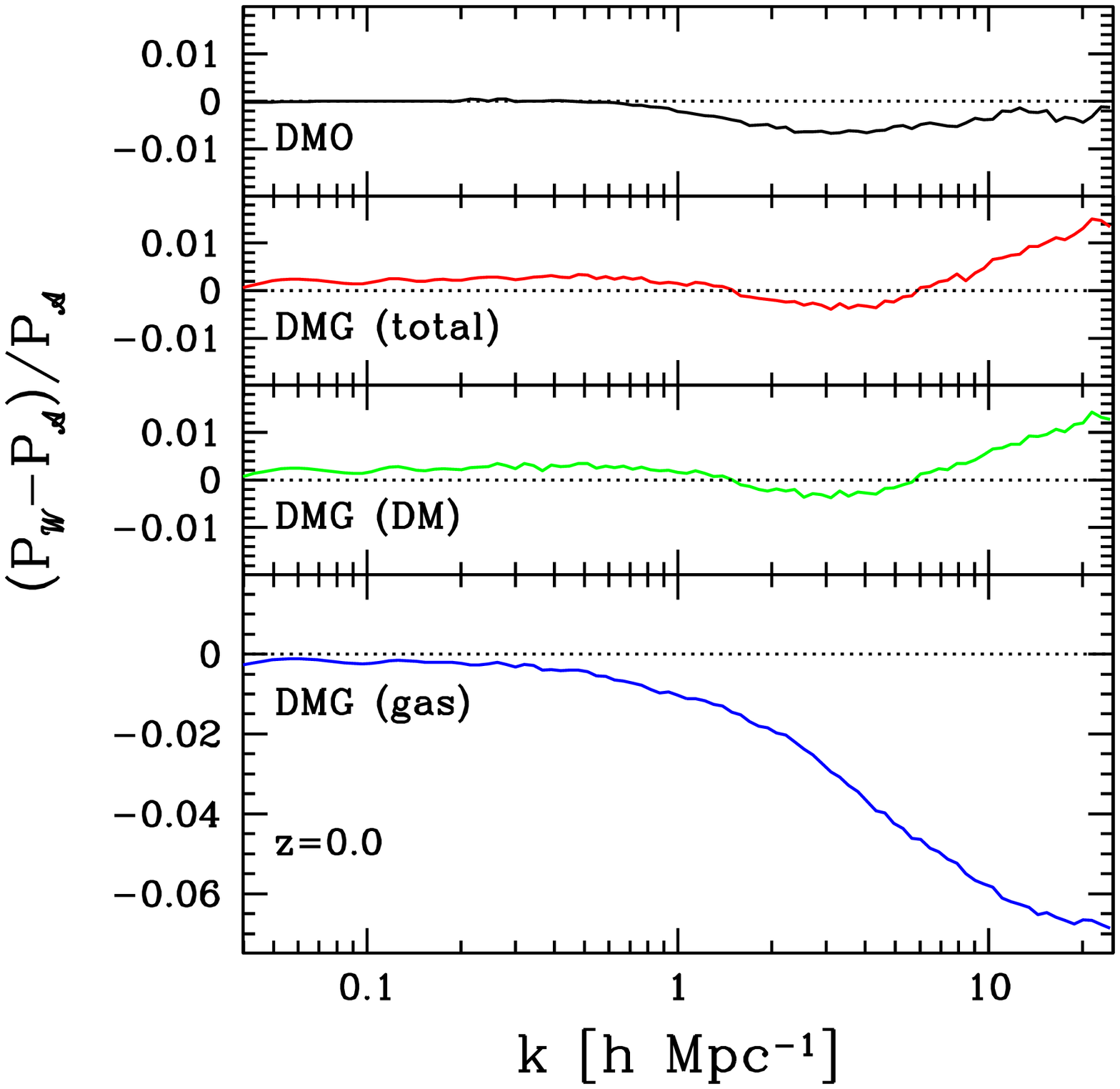,width=0.462\textwidth}
\psfig{figure=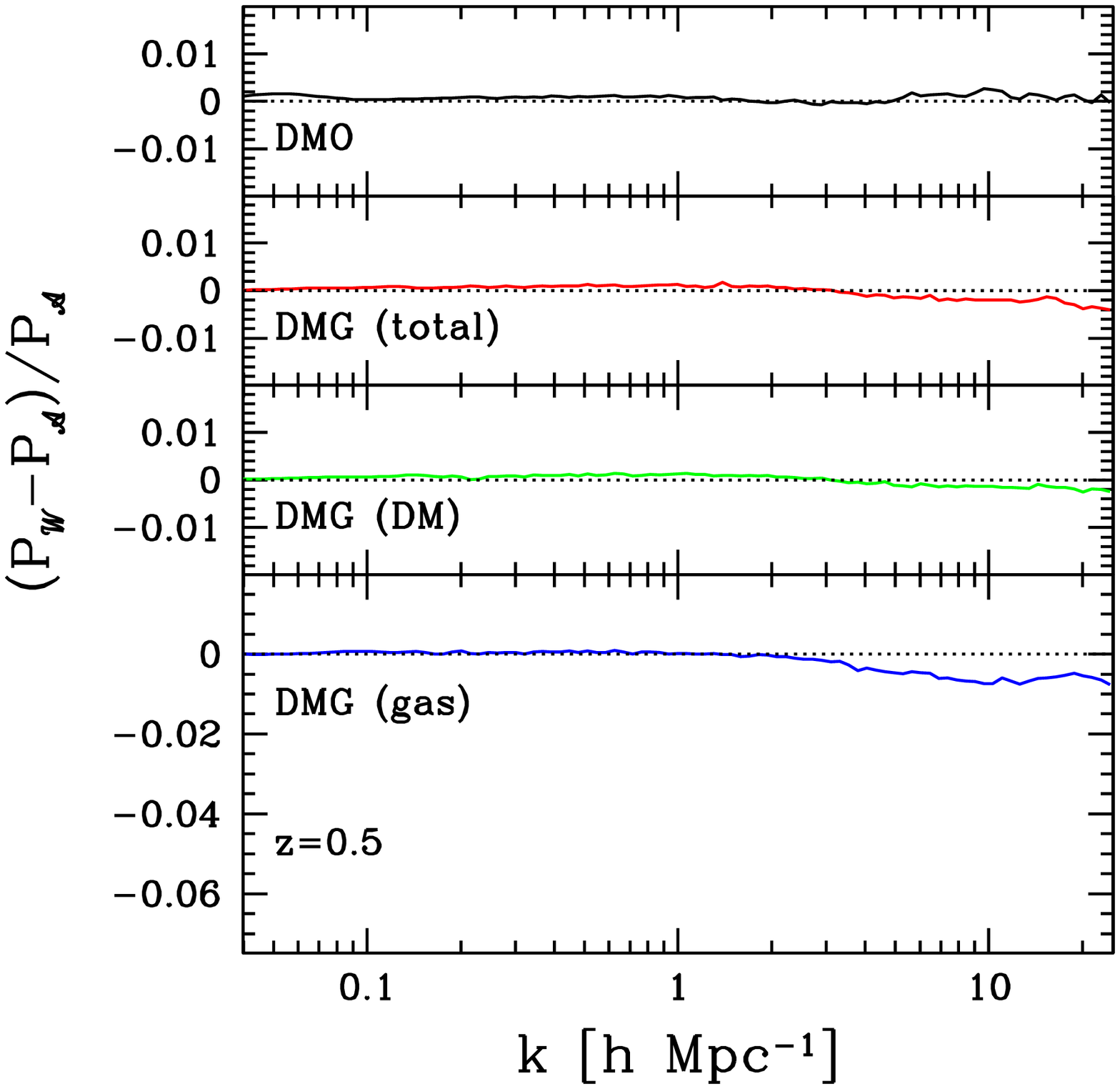,width=0.462\textwidth} 
\caption{Comparison between $\cal A$ and $\cal W$ spectra, for DMO and
the various components in DMG. Top and bottom panel show results for
$z=0$ and $z=0.5$ respectively.  }
\label{wr0}
\end{figure}

\begin{figure}
\psfig{figure=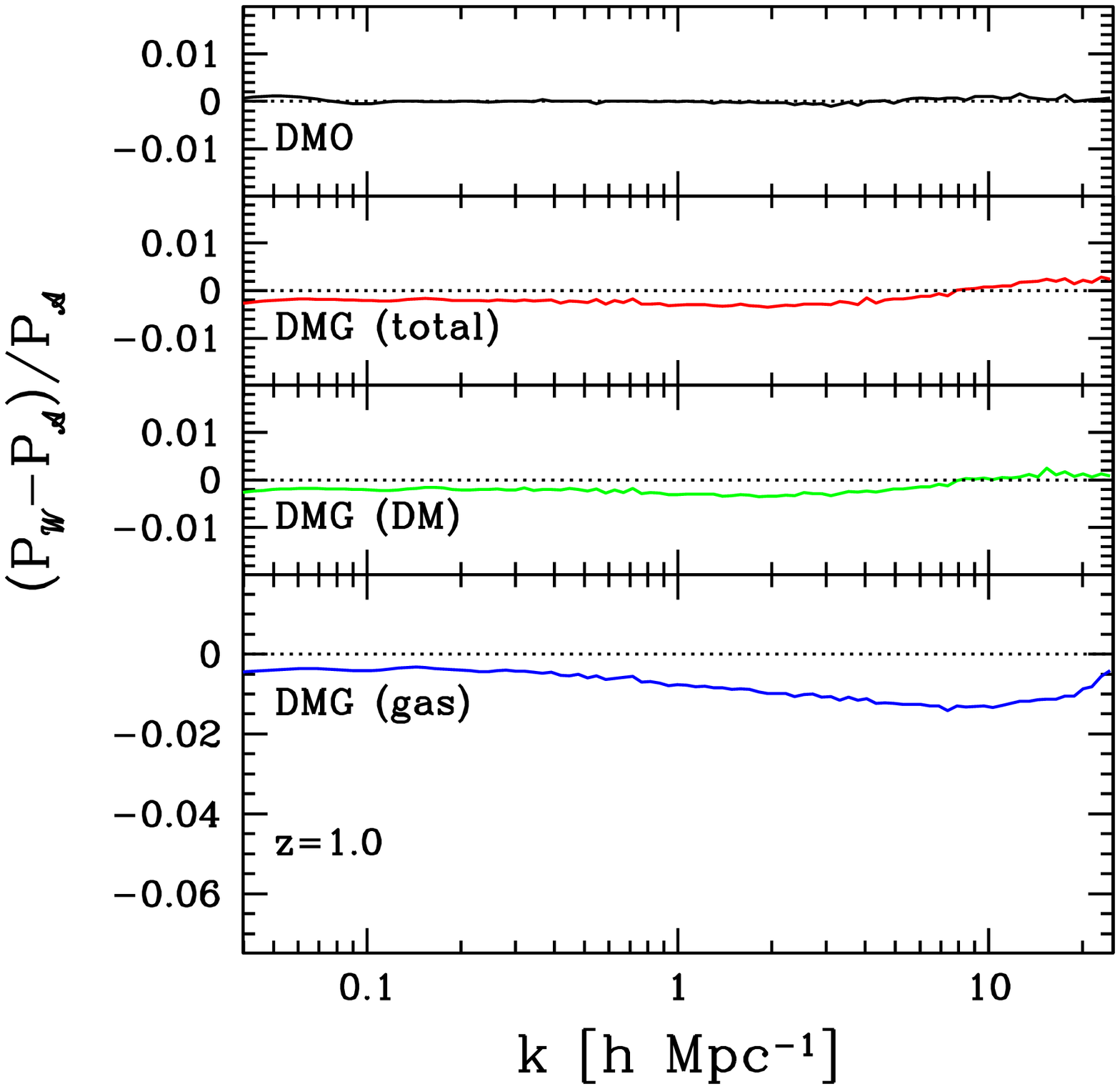,width=0.462\textwidth}
\psfig{figure=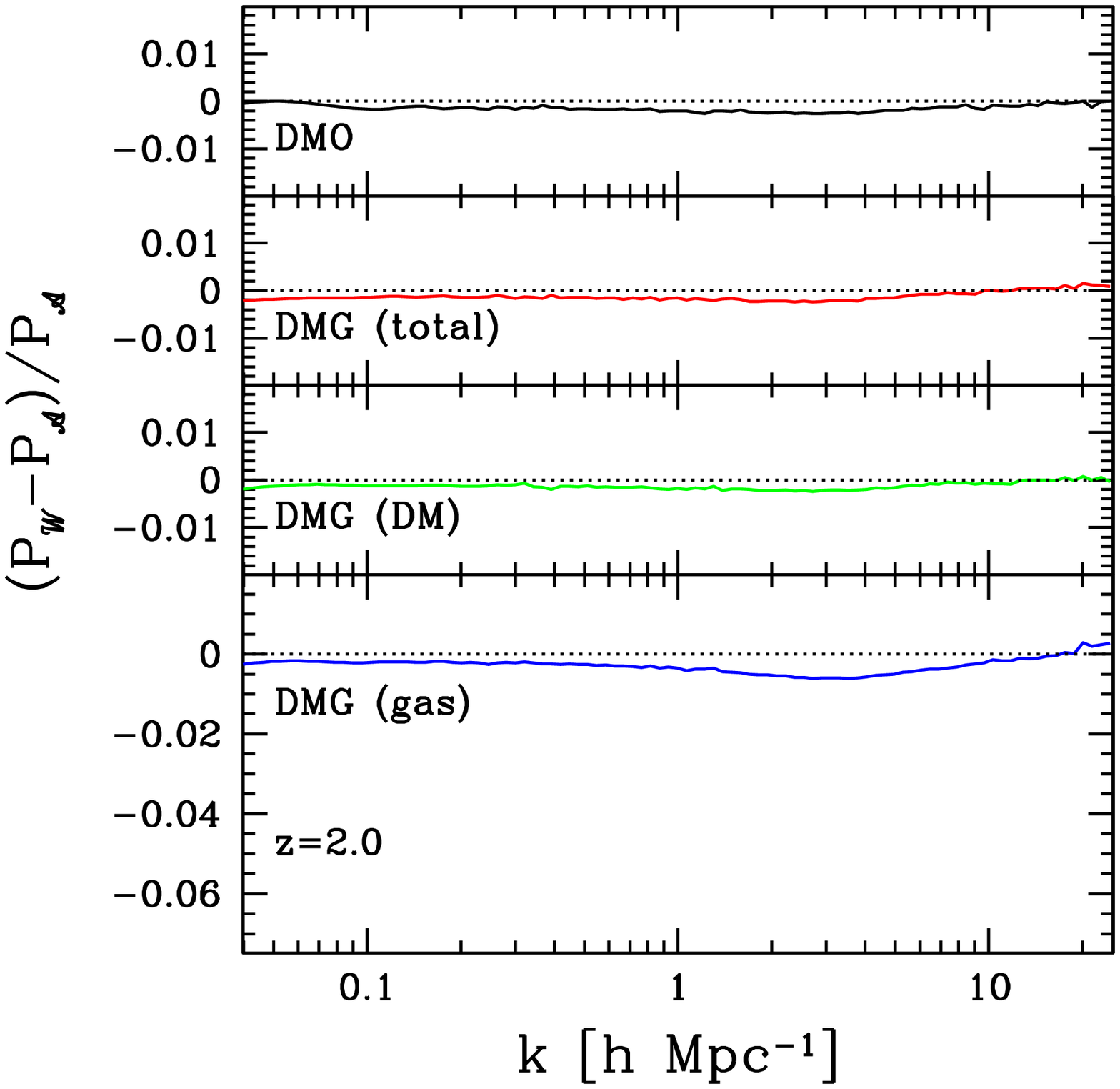,width=0.462\textwidth}
\caption{Same as figure \ref{wr0} but for $z=1$ (top panel) and $z=2$ (bottom panel).}
\label{wr0a}
\end{figure}

In Figures \ref{wr0} and \ref{wr0a} we compare spectra at $z = 0$ and 0.5, 1, 2~. 
Notice the extreme expansion of the ordinate units in these
plots.  Besides the gas case, the whole ordinate range is within $\pm
1 \, \%$, for $k<10 \Mpch$.

The basic result is that regularities persist when gas dynamics
is important, although residual discrepancies (of the order of a few
permils) are greater in DMG simulations. The largest discrepancy is
found in the gas spectra, even if the gas spectrum distortion at $z=0$,
attaining $\sim 5\, \%$, apparently does not imply a significant
distortion in the global spectrum.
Altogether it is fair concluding that even the inclusion of hydrodynamics
keeps the discrepancies between the auxiliary model and the true model in the permil range 
and this is one of of the major results of this work.

\subsection{Impact of  the results on next weak lensing data survey}

Matter density spectra $P(k,z)$ provide a basic link between observational data and theoretical models of the nature of DE.
We envisage comparing the models with data from
tomographic shear surveys.  Such surveys allow one to work out the angular
power spectrum $P^\kappa_{ij}(\ell)$, for weak lensing convergence
between the $i$--th and $j$--th tomographic beams covering the
redshift intervals $\Delta_{i}$ and $\Delta_{j}$. It is then easy to
show that $P^\kappa_{ij}(\ell)$ is directly related to $P(k,z),$ being
\begin{equation}
\label{PP}
P^\kappa_{ij}(\ell) = \left(H_o \over c \right)^3 
\int_0^\infty dz\, P\left[{\ell \over r(0,z)},z \right]
W_i(z) W_j(z){H_o \over H(z)}~.
\label{spectra}
\end{equation}
Here $H(z) $ is the Hubble parameter at the redshift $z$, $r(z,z')$ is
the comoving angular diameter distance between $z$ and $z'$, while
\begin{equation}
W_i(z) = 1.5\, \Omega_m (1+z) \int_{\Delta_i}
du ~ n_g(u)~r(u,z)/r(0,z)~,
\end{equation}
$n_g(z)$ being the comoving galaxy number distribution on redshift.
Using eq.~(\ref{spectra}), tomographic shear survey data, for $100 <
\ell < 3000$, will yield matter fluctuation spectra for $0.3 \sim <
k\, /h{\rm Mpc}^{-1} <\sim 10$.

The essential point, here, is that such spectra arise from any
possible gravitational source. Any link with assumptions on the
mass/luminosity ratio is therefore broken. Furthermore, the expected
accuracy of lensing surveys will allow to recover $P(k,z)$ with a
precision $\cal O$$(1\%)$.

For gravitational lensing shear predictions, at first order, there
seems to be no need to take into account the baryon component.
Baryons are comprise a small fraction of the Universal mass fraction
compared to DM, and their distribution can be only slightly different
from CDM. But what makes baryons extremely important is the precision
that shear measurements will reach ($\cal O$$(1\%)$) in the near
future. This presents a new challenge for hydrodynamical simulations.
Formerly, a slight inaccuracy in baryon physics was a second order
problem compared with the high precision achieved in tracing the total
matter distribution.  While baryons are less massive than CDM, the way
CDM reacts to the evolution of the baryon distribution makes baryons
an essential ingredient.

Let us then outline a very crucial issue concerning the
extrapolation of $w(z)$ from tomographic cosmic shear data.  This
point, already made in Paper I, is strengthened by this work, taking
into account baryon physics. The point is that the
conversion between the observed evolution of $w(z)$ and the true one
of the underlying model is not a straightforward operation.

\begin{figure}
\psfig{figure=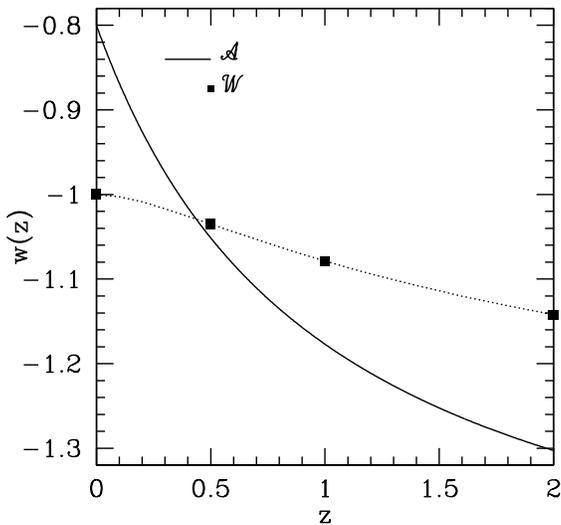,width=0.45\textwidth}
\caption{Evolution of the state parameter $w(z)$. The $\cal A$ model
is shown by the solid line and compared with the constant-$w$ values
of the auxiliary models $\cal W$ (dotted line) computed at different
redshifts. Solid squares show the $w$ values used in this work.}
\label{tw22}
\end{figure}

Figure \ref{tw22} shows the redshift evolution of the state parameter
of the $\cal A$ model (solid line) compared 
with the values of the constant-$w$ in the different auxiliary models
($\cal W$) at different redshifts (solid squares).
The figure outlines that
cosmic shear measures, fitted assuming constant $w$ in each
$z$--bin (i.e. the solid squares) would infer an evolution of $w$ with
redshift (the dotted line) radically different from the {\it true}
underlying model $\cal A$ (solid line).

The mapping between 
the $w$ values best fitting data in the tomographic layer about $z$
and the real evolution of $w(z)$ should be done with extreme care as
shown in Casarini (2010).

\section{Discussion and Conclusions}
\label{sec:dis}

In this work, we employed large scale hydrodynamical simulations, including gas cooling, 
star formation and Supernovae feedback.  Our simulations perform quite well when compared to 
observational results.
They produce a sizable stellar component in collapsed dark matter haloes
and star formation histories comparable with observations.
We are aware that these results strongly depend on the adopted (low) 
resolution, nevertheless the similarity to observations makes us confident that the effect of
baryons on the matter power spectrum in our simulation is sufficiently realistic. 
That said, further work is needed to better address the impact of the physics of star formation on the final 
results and, most important, to provide better tests of sample variance.
In particular, this work makes clear that no really quantitative
predictions, unaffected by sample variance, can be drawn from simulations, in $L\approx 50 \hMpc$ boxes.
Matching observed accuracies of
$\cal O$$(1\, \%)$ will require boxes a few hundred
Mpcs on a side. In principle a box with $L = 256\, h^{-1}$Mpc, as the one we
mostly used here, should be adequate to delve into haloes,
providing spectra up to $k \simeq 10\, h\, $Mpc$^{-1}$, with the
required accuracy. 
However, the spectra obtained from our large box differed in a non negligible way from the ones 
obtained from a smaller box ($L=64 \hMpc$). Although being fully compatible with the 
sample variance of the smaller box, this suggests further investigation will be required.

These residual uncertainties do not prevent us, though, from testing the
efficiency of a method introduced in Paper I, that allows us to work out
power spectra prediction for any DE model 
based on models with a constant equation of state ($w$).
We found up to $z = 2$, that the {\it auxiliary} $\cal W$
models (with constant $w$) yield spectra consistent with our {\it
assigned} $\cal A$ model (with a polynomial $w(z)$) within a few
permils. The test was performed for a single $\cal A$ dDE model. We
plan to extend such a test to more dDE models in forthcoming work. There
is however no reason to expect that the point selected on the
$w_o$--$w'$ plane owns peculiar properties.

The residual discrepancies between $\cal A$ and $\cal W$ models
exhibit a specific trend, already visible in Paper I:
discrepancies (of a few permils) are greater for smaller $z$ values. Also DMG results follow
the same rule, which is even even more pronounced for baryon spectra.
Gas spectra discrepancies often exceed the
permils but are systematically compensated by other components.
The success of our technique was not unexpected, but 
discrepancies at such a tiny level, also for $k > 2$--$3\, h\,
$Mpc$^{-1}$, exceeded our expectations.

The requirement that two models have the same
$\omega_{c,b}$ and $\sigma_8 (z)$ introduces  basic similarity: they
will have similar baryon acoustic oscillations and typical
fluctuation amplitudes.
Then, the requirement that $w$ yield the same conformal time $\tau$
elapsed since recombination for assigned and auxiliary
models apparently sets an equal timing for the gradual onset of
the non-linear evolution on analogous structures.
However, while $\tau$ is a fair time coordinate for structures 
slightly detached from the cosmic background, it no longer plays such
a role in cosmic islands, which have abandoned the cosmic expansion
and evolve on their own, forgetting the continuing increase of the
scale factor $a~.$ Within these structures, space-time is
substantially Minkowskian, and time evolution is set by the ordinary
time $t~.$

These are, however, two extreme situations. The scales which
``recently'' abandoned the overall expansion and entered a non-linear
regime are in an intermediate position. Our inspection mostly concerns
such scales. Accordingly, the success of an equal-$\tau$ requirement
is anything but granted.

Here we use hydrodynamical simulations to study a larger $k$ domain
(compared to Paper I), where the effects of baryons cannot be neglected anymore.
By introducing non-gravitational effects we expect dynamics to be
regulated by the ordinary time $t$, thus producing larger 
discrepancies at scales $k > 2$--$3\, h\, $Mpc$^{-1}$.
Figure \ref{wr0} rejects this hypothesis by showing that the matter
distribution is still ruled by the primeval clock.
In particular, for $k > 2$--$3\, h\, $Mpc$^{-1}$, the overall
DMG spectral discrepancies, at $z=0$, are perhaps even smaller than
DMO. Curiously enough, the opposite seems true at some greater $z$'s,
suggesting that most discrepancies found are pure noise, rather than
systematics. Clearly for larger $k$'s, when we delve into regions that
virialized long ago, $\cal A$ and $\cal W$ models unavoidably
differ. However, while we restrict ourselves to the scales that forthcoming weak lensing
surveys will study, the Paper I requirement still works unexpectedly well.


One ought to beware, however, that the success of our approach does
not imply that constant-$w$ and variable-$w$ models are
indiscernible.  What makes them observationally different is the
measured dependence of $w$ on the redshift $z$.

An essential issue, however, is that {\it the observed constant state parameter
$w(z)$ will NOT be the physical one of the DE component}. Although
we expect data to fit, at any $z$, the same density and Hubble
parameter values, the observed $w(z)$ will be the state parameter that
a constant-$w$ model must have, at such $z$, to grant the same
distance between there and the LSB.
The relation between the observed $w$ and the intrinsic one, for the DE component, 
should be put under control; otherwise there is a severe risk that future data
analysis lead to DE state equation misinterpretation.

A final point we wish to make is that {\sc halofit} expressions, so
useful for predicting spectra until now, have been tested and found
compatible with their claimed accuracy. Unfortunately, accuracy at
($\sim 6\, \%$) no longer matches our needs. Furthermore, among
models that approach the present data, {\sc halofit} expressions only need to handle
\LCDM.  If one aims to improve {\sc halofit} expressions, 
there is no longer any need to cover the wide array of models that once
were relevant, but now are far from observation.

Amelioration, however, is needed in two separate directions: (i) The
spectral range covered by predictions must reach $k \simeq 10\, h\,
$Mpc$^{-1}$, which requires that gas be considered; Rudd \etal (2008) had
actually attempted an extension in this direction, but limited themselves to
\LCDM cosmologies. (ii) DE state parameters different from constant $w
\equiv -1$ should be included.  Moreover, they should aim for a precision of
$\cal O$$(1\, \%)$, which presents a severe challenge, even for a much more
restricted model range.
Nevertheless when {\sc halofit} expressions of the required precision will be available,
even only for constant-$w$ models, thanks to the results of this work it will 
be straightforward to extend them to any  $w(z)$ law.

\section*{Acknowledgements} 

Giuseppe Murante is thanked for fruitful discussions. 
The authors acknowledge J. Wadsley for development of the {\sc gasoline} code 
and  thank him for its use in this work. Numerical
simulations and spectral analysis were performed on the PIA and on PanStarrs2 clusters of
the Max-Planck-Institut f\"ur Astronomie at the Rechenzentrum in
Garching and at the  CINECA--Bologna computer center, 
within the frame of the INAF--CINECA agreement (2008--2010).
The support of ASI (Italian Space Agency) through the
contract I/016/07/0 “COFIS” is acknowledged.


\end{document}